\def\Thetan{\Theta}

\font\fourteenbf=cmbx10 scaled\magstep2
\font\twelvebf=cmbx10 scaled\magstep1

%
%
%%%%%%%%%%%%%%%%%%%%%%%%%%%%%%%%%%%%%%%%%%%%%%%%%%%%%%%%%%%%%%%%%%%%%%%%
%\hoffset=1.0truein
%\voffset=1.0truein
%%%%%%%%%%%%%%%%%%%%%%%%%%%%%%%%%%%%%%%%%%%%%%%%%%%%%%%%%%%%%%%%%%%%%%%%
% preprint settings
%\hoffset= .5 true cm
%\vsize=22.5 true cm
%\hsize=16.0  true cm
%%%%%%%%%%%%%%%%%%%%%%%%%%%%%%%%%%%%%%%%%%%%%%%%%%%%%%%%%%%%%%%%%%%%%%%%
%\parskip=0.2cm
%\parindent=20pt
\raggedbottom
\baselineskip=16pt 
\overfullrule=0pt
%\font\poem=rpzcmi at 14truept
%\frac makes a nice fraction
\def\frac#1/#2{\leavevmode\kern.1em
 \raise.5ex\hbox{\the\scriptfont0 #1}\kern-.1em
 /\kern-.15em\lower.25ex\hbox{\the\scriptfont0 #2}}
%%%%%%%%%%%%%%%%%%%%%%%%%%%%%%%%%%%%%%%%%%%%%%%%%%%%%%%%%%%%%%%%%%%%%%%%
% submission settings
% \magnification=\magstep1
% \baselineskip=20pt
%%%%%%%%%%%%%%%%%%%%%%%%%%%%%%%%%%%%%%%%%%%%%%%%%%%%%%%%%%%%%%%%%%%%%%%%
\def\approx{\simeq}
%%%%%%%%%%%%%%%%%%%%%%%%%%%%%%%%%%%%%%%%%%%%%%%%%%%%%%%%%%%%%%%%%%%%%%%%
%
%   Here come macros for equation numbering.
%
\catcode`@=11
\newcount\chapternumber      \chapternumber=0
\newcount\sectionnumber      \sectionnumber=0
\newcount\equanumber         \equanumber=0
\let\chapterlabel=0
\newtoks\chapterstyle        \chapterstyle={\Number}
\newskip\chapterskip         \chapterskip=\bigskipamount
\newskip\sectionskip         \sectionskip=\medskipamount
\newskip\headskip            \headskip=8pt plus 3pt minus 3pt
\newdimen\chapterminspace    \chapterminspace=15pc
\newdimen\sectionminspace    \sectionminspace=10pc
\newdimen\referenceminspace  \referenceminspace=25pc
\def\chapterreset{\global\advance\chapternumber by 1
   \ifnum\the\equanumber<0 \else\global\equanumber=0\fi
   \sectionnumber=0 \makel@bel}
\def\makel@bel{\xdef\chapterlabel{%
\the\chapterstyle{\the\chapternumber}.}}
\def\sectionlabel{\number\sectionnumber \quad }
\def\unnumberedchapters{\let\makel@bel=\relax \let\chapterlabel=\relax
\let\sectionlabel=\relax \equanumber=-1 }
\def\eqname#1{\relax \ifnum\the\equanumber<0
     \xdef#1{{\rm(\number-\equanumber)}}\global\advance\equanumber by -1
    \else \global\advance\equanumber by 1
      \xdef#1{{\rm(\chapterlabel \number\equanumber)}} \fi}
\def\eqinsert#1{\noalign{\dimen@=\prevdepth \nointerlineskip
   \setbox0=\hbox to\displaywidth{\hfil #1}
   \vbox to 0pt{\vss\hbox{$\!\box0\!$}\kern-0.5\baselineskip}
   \prevdepth=\dimen@}}
%

%

%

%
%%%%%%%%%%%%%%%%%%%%%%%%%%%%%%%%%%%%%%%%%%%%%%%%%%%%%%%%%%%%%%%%%%%
%
\newcount\fcount \fcount=0
\def\ref#1{\global\advance\fcount by 1 
  \global\xdef#1{\relax\the\fcount}}
\def\refs{\noindent \hangindent=5ex\hangafter=1}

%

% \np starts a new page
%
%  macro for invoking today's date (when TeX is run on your file)
%
\def\today{\number\year\space \ifcase\month\or 	January\or February\or 
	March\or April\or May\or June\or July\or August\or September\or
	October\or November\or December\fi\space \number\day}
%
% Some useful symbols
%
%\simlt and \simgt produce > and < signs with twiddle underneath
\def\spose#1{\hbox to 0pt{#1\hss}}
\def\simlt{\mathrel{\spose{\lower 3pt\hbox{$\mathchar"218$}}
     \raise 2.0pt\hbox{$\mathchar"13C$}}}
\def\simgt{\mathrel{\spose{\lower 3pt\hbox{$\mathchar"218$}}
     \raise 2.0pt\hbox{$\mathchar"13E$}}}
%\simpropto produces \propto with twiddle underneath
\def\simpropto{\mathrel{\spose{\lower 3pt\hbox{$\mathchar"218$}}
     \raise 2.0pt\hbox{$\propto$}}}

\unnumberedchapters
 
% JUNK FOR NICE COVER PAGE

 at 14.4truept
%\font\titlefont=rptmb at 12truept
\font\namefont=cmr12 
\font\addrfont=cmti12 

% ABSTRACTS:
\newbox\abstr
\def\abstract#1{\setbox\abstr=
    \vbox{\hsize 5.0truein{\par\noindent#1}}
    \centerline{ABSTRACT} \vskip12pt 
    \hbox to \hsize{\hfill\box\abstr\hfill}}

%  macro for invoking today's date (when TeX is run on your file)
\def\today{\ifcase\month\or
        January\or February\or March\or April\or May\or June\or
        July\or August\or September\or October\or November\or December\fi
        \space\number\day, \number\year}

\def\author#1{{\namefont\centerline{#1}}}
\def\addr#1{{\addrfont\centerline{#1}}}

\def\ie {{\it i.e.}}
\def\eg {{\it e.g.}}
\def\etal {{\it et al.}}
{ % SETTINGS FOR COVER PAGE ONLY
{\nopagenumbers
{
\vsize=9 truein
\hsize=6.5 truein
\raggedbottom
\baselineskip=18pt
%\today
\hfill IASSNS-AST-96/30 %, CfPA-TH-95-09, UTAP-204, astroph-9504057
 
\hfill 
\vskip3truecm
\centerline {
 \twelvebf  
{\fourteenbf T}HE
{\fourteenbf P}HYSICS 
{\fourteenbf O}F } 
\centerline {
 \twelvebf  
{\fourteenbf M}ICROWAVE 
{\fourteenbf B}ACKGROUND
{\fourteenbf A}NISOTROPIES
\footnote{$^{\rm\dag}$}{\rm\negthinspace\negthinspace\negthinspace
Review for Nature, revised \today.} 
            }
\nobreak
\baselineskip=15pt
  \vskip 0.5truecm
  \author{Wayne Hu,$^1$ Naoshi Sugiyama,$^{2}$ \& Joseph Silk$^{3}$}
  \smallskip
  \addr{$^{1}$Institute for Advanced Study}
  \addr{School of Natural Sciences}
  \addr{Princeton, NJ 08540}
  \smallskip
  \addr{$^{2}$Department of Physics}
  \addr{and Research Center for the Early Universe}
  \addr{The University of Tokyo, Tokyo 113, Japan}
  \smallskip
  \addr{$^{3}$Department of Astronomy and Physics}
  \addr{and Center for Particle Astrophysics}
  \smallskip
  \bigskip
  \bigskip	 
%\abstract{ %\baselineskip=15pt \rm 
\noindent{\rm 
Cosmic microwave background anisotropies   
provide a vast amount of cosmological information.  Their full physical 
content and detailed structure
can be understood in a simple and intuitive fashion
through a systematic investigation of the individual mechanisms for
anisotropy formation.
\smallskip
%{\noindent\it 
%Subject Headings: cosmic microwave background - large scale structure
%of universe}
}}
\bigskip
\vfill 
\noindent whu@sns.ias.edu

\noindent Additional material can be
found at {\tt http://www.sns.ias.edu/$\sim$whu/physics/physics.html}
\eject}}
\bigskip 
%\bigskip
%\noindent{\twelvebf Introduction}
%\smallskip
\noindent 

\ref\SSW
\ref\Brandt
\ref\TegEfs
\ref\Smoot
\ref\PY
\ref\SZ
\ref\SW
\ref\Silk
\ref\Bardeen
\ref\Mukhanov
\ref\KS
\ref\WSS
\ref\Bond
\ref\HSa
\ref\DZS 
\ref\HSb
\ref\HW
\ref\SG
\ref\KSS
\ref\Kofman
\ref\SS
\ref\RS
\ref\SelRS
\ref\TL
%\ref\DNP
\ref\AW 
\ref\Star
%\ref\Crittenden
\ref\Davis
\ref\KT
%\ref\BSB
\ref\Coulson
\ref\PST
\ref\BS
\ref\CE
\ref\SeljakGL
\ref\Kaiser
\ref\OV
\ref\Vishniac
\ref\ADPG
\ref\SZkin
%\ref\HWvish
%\ref\DDP
%\ref\PSCO
%\ref\HT
%\ref\CB
%\ref\KSY
\ref\CT
\ref\Albrecht
\ref\WS
\ref\BE
\ref\VS
\ref\HSSW
\ref\MB
\ref\Confusion
\ref\Seljak
\ref\Jungman
%\ref\LS
\ref\Bucher
\ref\YST
\ref\DGS

An extraordinary wealth of cosmological information lies in the
cosmic microwave background (CMB) temperature anisotropy at degree 
and subdegree scales.  Yet despite the flurry of reported measurements 
in the past several years,$^\SSW$ 
systematic errors, extragalactic sources 
and galactic 
foregrounds have prevented definitive measurements
at these scales.  
With adequate sky and 
frequency coverage, foreground contamination can be
avoided or removed.$^{\Brandt,\TegEfs}$  
Experiments   
presently in the planning stage are expected to overcome such
difficulties within the next few years.  
It is therefore timely to
review the physics of anisotropy formation in order to lay the 
basis for the science that may be extracted from these anticipated
detections.  

A definitive set of experiments
must achieve good sensitivity at an
an angular resolution of at least 10 arcminutes.  
%It must also attain
%adequate sky and frequency coverage in order to avoid or remove
%various sources of foreground contamination.$^{\Brandt}$
Large angle anisotropies, detected by the {\it COBE} satellite,$^\Smoot$
probe the
fluctuations laid down in the very early universe, possibly
the result of quantum mechanical processes during an epoch of
inflation.  These primordial fluctuations grew by gravitational
instability into the large scale structure of the universe today. 
To reveal more cosmological information, one has to probe
the anisotropy of the CMB at smaller angular scales where causal
interactions take place.  A causally-connected region, or horizon volume,
at last scattering subtends no more than a degree on the sky under
most conditions. 

Inside the horizon, acoustic,$^\PY$ Doppler,$^\SZ$ gravitational
redshift,$^\SW$ and photon diffusion$^\Silk$ 
effects combine to form 
a seemingly complicated spectrum of primary anisotropies.
Considering the component contributions individually 
reveals the underlying simplicity and sensitivity of the spectrum
to a variety of cosmological
parameters, including the baryon density, the dark matter density,
the cosmological constant, the Hubble constant, and the curvature of the
universe. 
Secondary effects generated at later times may provide 
important clues for 
the process of structure formation.
Furthermore, these degree and subdegree scale anisotropies 
will enable one to attempt a reconstruction of the spectrum 
and evolution of density fluctuations on 100 Mpc scales which 
deep galaxy redshift surveys are beginning 
to probe.

In this review, we draw on classical Newtonian analogues to 
represent anisotropy formation.$^{\Bardeen,\Mukhanov}$  
We refer the reader elsewhere 
for alternate descriptions, which differ in perspective
not in physical content, and more complete historical 
developments of the subject.$^{\KS,\WSS,\Bond}$  

\bigskip 

\noindent{\twelvebf Primary Anisotropies}
\smallskip
\noindent 

Before redshift $z_* \approx 10^3$, CMB photons are hot enough to 
ionize hydrogen.  Compton scattering tightly couples the
photons to the electrons which are in turn coupled to the baryons
by electromagnetic interactions.  The system can thus be dynamically
described as a photon-baryon fluid. Photon pressure resists
gravitational compression of the fluid and sets up acoustic oscillations. 
At $z_*$, neutral hydrogen forms and the photons last scatter.  Regions
of compression and rarefaction at this epoch represent hot and cold spots
respectively.  Photons also suffer gravitational 
redshifts from climbing out of the potentials on the last scattering
surface.  The resultant fluctuations 
appear to the observer today as anisotropies on the sky.   
We call these fluctuations {\it primary} anisotropies.  Secondary
anisotropies can also be generated between recombination and the
present.  We consider those contributions in the next section.

\goodbreak\smallskip
\noindent{\bf Normal Modes}

\def\meff{m_{\rm eff}} 
\noindent
Normal mode analysis breaks the system into independent oscillators.
In flat space, this corresponds to a Fourier decomposition of the
fluctuation into plane waves of comoving wavenumber $k$.  
An isotropic temperature perturbation 
$\Thetan = \Delta T/T$ in mode $k$ evolves almost as a  
simple harmonic oscillator before recombination$,^{\HSa}$
$\meff \ddot \Thetan + k^2 c^2 \Theta/ 3 \approx \meff g$.  
The overdots represent derivatives with respect to conformal 
time $\eta = \int (1 + z) dt$ and the effective dimensionless 
mass of the oscillator is $\meff = 1+R$.  Here 
$R=(\rho_b+p_b)/(\rho_\gamma+p_\gamma) \approx 
3\rho_b/4\rho_\gamma$ % = 3.0 \times 10^{4} (1+z)^{-1} \Omega_B h^2$ 
is the baryon-photon momentum density ratio.
The oscillation frequency obeys the dispersion 
relation $\omega = kc/ \sqrt{3 \meff} = kc_s$, where
$c_s$ is the sound speed.
Gravity provides an effective 
acceleration of $g = - k^2 c^2 \Psi/3 -\ddot \Phi$, where $\Psi$
is the Newtonian gravitational potential
and $\Phi \approx -\Psi$ is the curvature perturbation on spatial
hypersurfaces.
These are related to the density fluctuation via the Poisson equation
and give rise to gravitational infall and time dilation as we shall
see.

\goodbreak\smallskip
\noindent{\bf Acoustic Oscillations}

\noindent Let 
us first consider 
the case of a {\it static} potential.$^{\DZS}$
Here, $g \approx -k^2 c^2\Psi /3$ and supplies
the usual gravitational force that causes matter to fall
into potential wells.
Since big bang nucleosynthesis implies that the baryon
density is low, as a first approximation,
assume that the photons completely 
dominate the fluid, $R \ll 1$. % \meff=1$.  % and $c_s \approx c/\sqrt{3}$. 

\input epsf.tex
\topinsert
\centerline{
\epsfxsize=6.50in \epsfbox{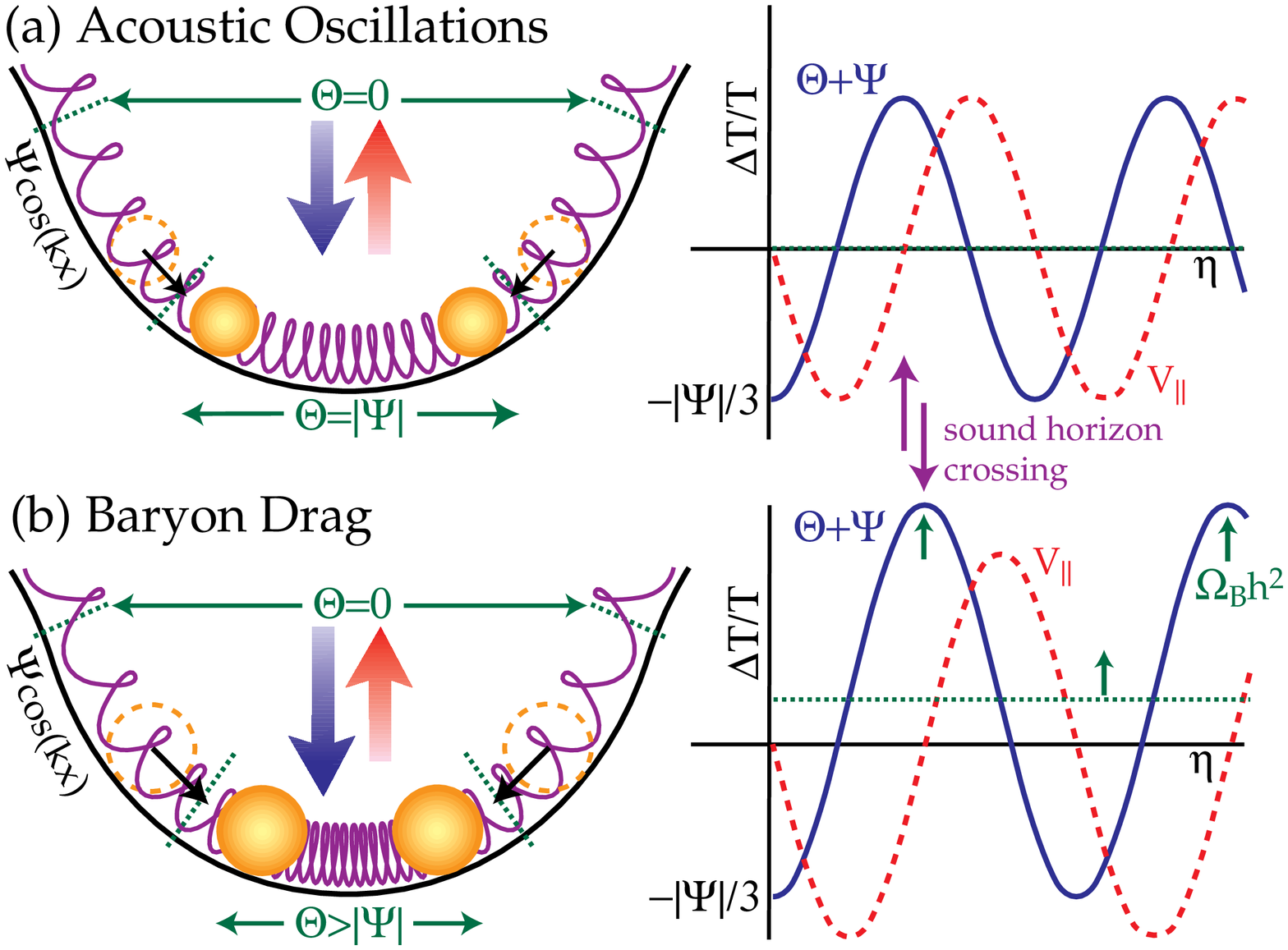}}

\baselineskip=12truept \rightskip=3truepc \leftskip=3truepc
\noindent {\bf Figure 1.} (a) Acoustic oscillations.  
Photon pressure resists gravitational compression of the fluid
setting up acoustic oscillations (left panel, real space 
$-\pi/2 \simlt kx \simlt \pi/2$).  Springs and balls schematically
represent fluid pressure and effective
mass respectively. Gravity displaces
the zero point such that $\Theta\cos(kx) = -\Psi\cos(kx)$ at 
equilibrium with oscillations in time of amplitude $\Psi/3$ (right
panel).
The displacement is cancelled by the redshift $\Psi\cos(kx)$ a photon
experiences climbing out of the well.  Velocity 
oscillations lead to a Doppler effect $V_\parallel$ shifted by 
$\pi/2$ in phase from the temperature perturbation.  (b) Baryon drag
increases the gravitating mass, causing more infall and a net 
zero point displacement, even after redshift.  Temperature
crests (compression) are enhanced over troughs (rarefaction) 
and Doppler contributions. 

\endinsert

Gravitational infall compresses the fluid until resistance from
photon pressure reverses the motion. 
To complete the description, we 
need to choose the initial conditions, assumed for now to be 
adiabatic (see Box 1):
$\Thetan(0)=-{2 \over 3} \Psi$ and $\dot \Thetan(0)=0$.
Since a constant gravitational force merely shifts
the zero point of the oscillation to $\Thetan= -\Psi$,
the initial displacement of  
${1\over3}\Psi$ 
from the zero point 
enters into the oscillation as $\Thetan  + \Psi
= {1 \over 3} \Psi\cos(ks)$, where the {\it sound horizon} is
$s = \int c_s d\eta$ (see Fig.~1a).
At last scattering $\eta_*$, 
the photons
decouple from the baryons and stream out of potential wells suffering
gravitational redshifts equal to $\Psi$.  
%They then suffer gravitational redshifts equal to $\Psi$ 
%as they climb out of the 
%potential well making the resulting fluctuation $\Thetan(\eta_*)+\Psi 
%= {1 \over 3} \Psi\cos(kc_s\eta)$.
We thus call $\Thetan+\Psi$ the {\it effective} temperature
fluctuation.
Here the 
redshift exactly cancels the zero
point displacement since gravitational
infall and redshift have the same physical origin if the
baryons are dynamically insignificant. 

The phase of the oscillation is frozen in at last scattering. 
The critical wavenumber $k_A = \pi /s_*$ 
corresponds to the sound horizon at that time (see Fig.~1a). 
Longer wavelengths
will not have evolved from the initial conditions and possess
${1 \over 3} \Psi$ fluctuations after gravitational 
redshift. This combination of the intrinsic 
temperature fluctuation and the gravitational redshift is the 
well-known Sachs-Wolfe effect$.^{\SW}$  Shorter wavelength fluctuations 
can be frozen
at different phases of the $\cos(ks_*)$ oscillation.  
As a function of $k$, there will be a harmonic
series
of temperature {\it fluctuation} 
peaks with $k_m = mk_A = m\pi/s_*$ for the $m$th peak.
Odd peaks thus represent the compression phase (temperature crests),
whereas even peaks represent the rarefaction phase (temperature troughs), 
inside the potential wells.  More exotic models might produce
a phase shift $\cos(ks_* + \phi)$, but even so the spacing
between the peaks remains $k_m - k_{m-1} = k_A$ (see Box 1).
Thus the sound horizon at last scattering  $s_*(\Omega_0 h^2,\Omega_B h^2)$
should be measurable from the CMB.
Here the Hubble constant is $H_0=100 h$ km s$^{-1}$
Mpc$^{-1}$ and $\Omega_0$ and $\Omega_B$ are the present matter 
and baryon density in units of the critical density.  The
dependence on $\Omega_Bh^2$ is weak if $R(z_*) \approx 30\Omega_Bh^2 
\ll 1$.

\goodbreak\smallskip
\noindent{\bf Baryon Drag}

\noindent
While effectively pressureless, the baryons do
contribute to the mass of the fluid $m_{\rm eff} = 1+R$, where
recall $R \propto \Omega_B h^2$. 
This changes the
balance between pressure and gravity. 
In the presence of baryons, gravitational infall 
leads to greater compression of the 
fluid in a potential
well, \ie\ a further
displacement of the oscillation zero point$^{\HSa}$ (see Fig.~1b).  
Since the redshift is not affected
by the baryon content, this relative shift remains after 
last scattering to enhance all peaks from compression over those
from rarefaction.
If $R$ were constant, 
$\Thetan+\Psi = {1 \over 3}\Psi (1+3R)\cos(ks) -R\Psi$,
with 
compressional peaks a factor of $(1+6R)$ larger than the
Sachs-Wolfe plateau and a difference in peak amplitude of 
$2R\Psi$ between even
and odd
\ref\HSsmall peaks.$^{\HSsmall}$
In reality, these effects are reduced since 
$R\rightarrow 0$ at early times.  Nevertheless, the relative peak heights
probe $\Omega_B h^2$ through $R$ and the amplitude of
potential perturbations at last scattering through $\Psi$.

Finally the {\it evolution} of the effective mass influences 
the oscillation.
In classical mechanics, the 
ratio of energy ${1 \over 2} m_{\rm eff} \omega^2
A^2$ to frequency $\omega$ of 
an oscillator is an adiabatic 
invariant.  Thus for the slow changes in 
$m_{\rm eff} \propto \omega^{-2}$,
the amplitude of the oscillation varies as $A \propto m_{\rm eff}^{-1/4} 
\propto (1+R)^{-1/4}$ representing a small decay with time.  
%Since $R(z_*) = 30\Omega_B h^2
%\simlt 1$ at recombination, this is ordinarily not a strong effect.
 
\noindent{\bf Doppler Effect}

\noindent
Since the turning points are at
the extrema, the fluid velocity oscillates $\pi/2$
out of phase with the density (see Fig.~1a).  
Its line-of-sight motion relative to the observer causes a 
Doppler shift.  
Whereas the observer velocity creates a pure dipole anisotropy
on the sky, the fluid velocity $v_\gamma$
causes an rms spatial temperature variation 
$V_\parallel = {1 \over \sqrt{3}} v_\gamma/c$ 
on the last scattering surface from its line-of-sight component.$^{\SZ}$ 
For a photon-dominated $m_{\rm eff} \approx 1$ fluid,
the velocity contribution is equal in amplitude to the density 
effect$^{\DZS}$ $V_\parallel = {1 \over 3}\Psi \sin(ks)$. 
%$.^{\DSZ,\Jorg}$ 

The addition of baryons changes
the relative velocity contribution.  As the effective mass
increases, conservation of energy requires that
the velocity decreases for the same initial temperature displacement. 
Thus the {\it relative} amplitude of the Doppler effect scales as 
$m_{\rm eff}^{-1/2}$.  
In the toy model of a constant baryon-photon momentum ratio $R$, the 
contribution
becomes $V_\parallel = {1 \over 3}\Psi (1+3R)(1+R)^{-1/2} 
\sin(ks)$. 
Notice that velocity oscillations are symmetric around zero leading to
an even more prominent compressional peaks (see Fig.~1a).
Even in a universe with $\Omega_B h^2 \approx 10^{-2}$ given 
by nucleosynthesis, $R$ is sufficiently large
to make velocity contributions subdominant. 

\goodbreak\smallskip
\noindent{\bf Driving Effect}

\noindent 
All realistic models involve time-dependent potentials.
Forced acoustic oscillations result and can
greatly enhance the peaks if the forcing frequency 
matches the natural frequency.  Such is the case if the
acoustic perturbations themselves generate most of the
force through their self-gravity.$^{\HSb,\HW}$  

We have hitherto assumed that matter dominates the energy density. 
In reality,
radiation dominates 
above the redshift of equality 
$z_{\rm eq}= 2.4 \times 10^4 \Omega_0 h^2$, assuming the
usual three flavors of massless neutrinos.   
We show in Box 1 how radiation fluctuations introduce a distinct
class of {\it isocurvature} perturbations and distinguishes them
from adiabatic ones by the associated driving mechanism.
In both cases, the driving effect boosts the amplitude
of the oscillations and leaves an imprint of the matter-radiation
transition in the CMB. The critical scale 
corresponds to the wavenumber crossing the horizon at $z_{\rm eq}$,
$k_{\rm eq} = 7.3 \times 10^{-2} \Omega_0 h^2 $ Mpc$^{-1}$.  
For adiabatic fluctuations, acoustic
oscillations smaller than this scale 
are boosted by a factor of $\sim 5$ above
the Sachs-Wolfe plateau in temperature.  
The peak-to-plateau ratio thus probes
$k_{\rm eq}$ and so $\Omega_0 h^2$.

\noindent{\bf Photon Diffusion}

\noindent In reality, coupling is imperfect since the photons 
possess a mean free path to Compton scattering $\lambda_C$. 
As the photons random walk through the baryons, hot and cold
regions are mixed.  Fluctuations damp
nearly exponentially 
as the diffusion length
$\lambda_D \sim \sqrt{N}\lambda_C = \sqrt{c\eta\lambda_C}$ overtakes
the wavelength$.^\Silk$

At last scattering, the ionization fraction $x_e$ 
decreases due to recombination, 
thus
increasing the mean free path of the photons $\lambda_C \propto
(x_e n_b)^{-1}$. 
The effective diffusion scale $k_D$ can therefore
be used as a probe of the ionization history and the baryon 
content $\Omega_B h^2$.  Notice that this scale is not dependent on
the nature of the initial fluctuations assumed ({\it cf.}~Box 1).
If recombination is delayed, \eg\ by
the early release of energy from particle decays or non-linear
fluctuations, diffusion continues. 
As the diffusion length reaches the sound horizon, the
acoustic peaks vanish.  This occurs in
models where recombination never occurred.  We will return, below,
to consider
the less extreme variant of recombination followed by 
late reionization.

\goodbreak\smallskip
 
\topinsert
\centerline{
\epsfxsize=5.75in \epsfbox{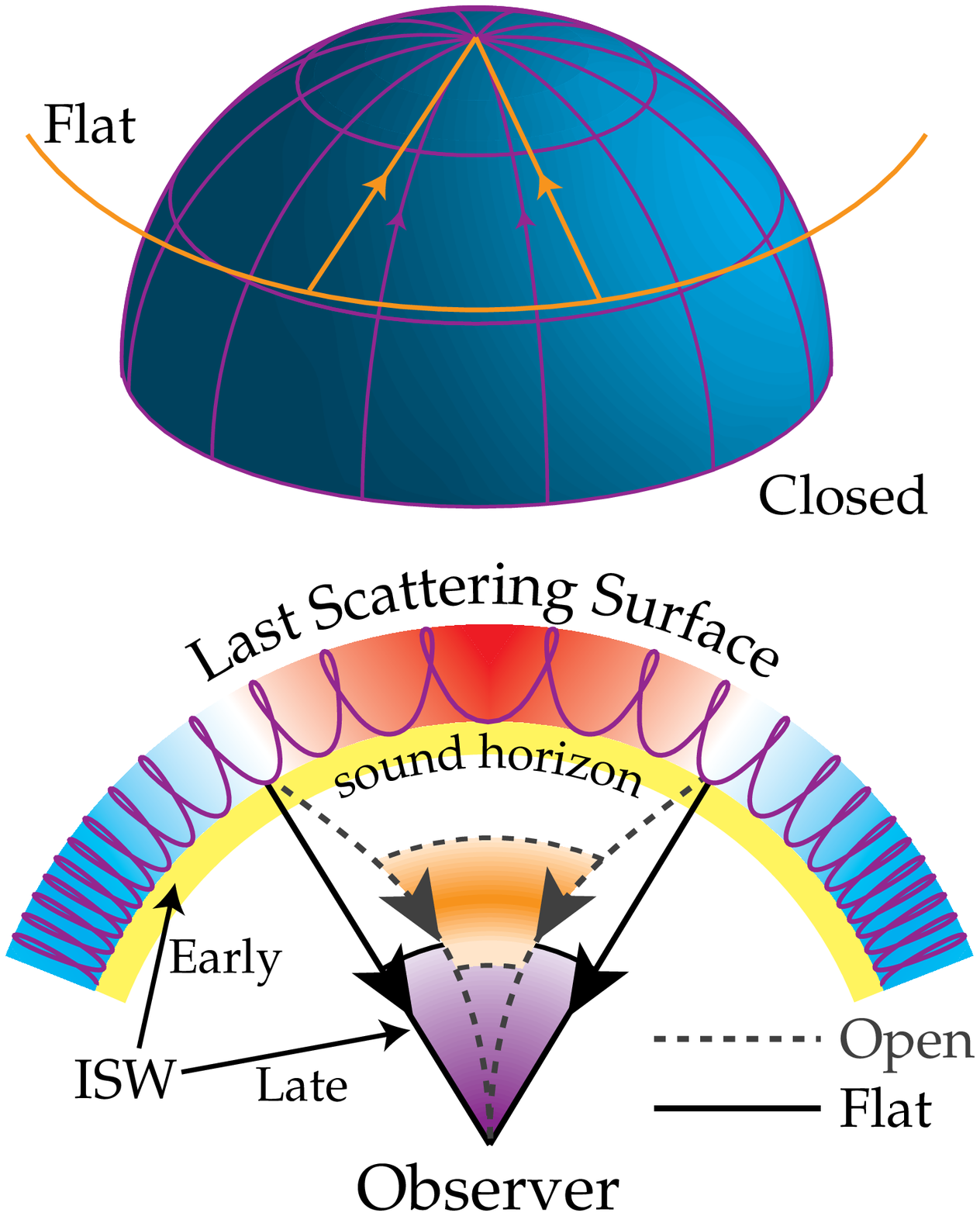}}

\baselineskip=12truept \rightskip=3truepc \leftskip=3truepc
\noindent {\bf Figure 2.}  Projection effects.  Temperature fluctuations
on a distant surface appear as anisotropies on the sky.  The angular
size depends on the geometry of 
the universe and the distance to this surface.  
At a fixed distance, a smaller physical scale is required to subtend the
same angle in a closed universe and larger in an open universe 
(schematically flattened for clarity).
Acoustic fluctuations from last scattering
subtend a smaller angle on the sky than the ISW effects for the
same physical scale.  

\endinsert

\smallskip
\noindent{\bf Projection Effect}

\noindent The description of the primary signal now lacks only the
relation between fluctuations at $z \approx 10^3$ and 
anisotropies today.
A spatial fluctuation
on this distant surface appears as an 
anisotropy on the sky.  
Two quantities affect the projection: the curvature of the universe
and the distance to the surface.
The curvature is defined as $K = - H_0^2 \Omega_K /c^2$.  Here the 
relative contribution of the curvature to the expansion rate is
$\Omega_K = 1 - \Omega_0 - \Omega_\Lambda$, with $\Omega_\Lambda$
related to the cosmological constant as 
$\Lambda = 3H_0^2 \Omega_\Lambda/c^2$.

Consider first the case
of positive curvature.  Photons free stream to the observer on
geodesics analogous to lines of longitude to the pole (see Fig.~2).  
Positive
curvature (closed universe $K > 0$)  
makes the same physical scale at fixed latitude (distance)
subtend a larger angle than in the Euclidean case.  The opposite 
effect occurs for negative curvature$^{\DZS,\SG}$ 
(open universe $K < 0$). 

Increasing the distance
to the last scattering surface also decreases the angular extent of
the features.  The distance to last scattering $c(\eta_0-\eta_*)$  
depends mainly on the
expansion rate and hence on $H_0$,$\Omega_0$,$\Omega_K$ 
and 
very weakly on $\Omega_\Lambda$.
Putting these quantities together gives us the angular
diameter distance $d = |K|^{-1/2} \sinh[|K|^{1/2}c(\eta_0-\eta_*)]$
for $K<0$ and similarly for $K>0$ with $\sinh \rightarrow \sin$.
The angular extent $\theta \sim \ell^{-1}$ of a physical feature
in the CMB is given by  
$\ell_{\rm feature} = k_{\rm feature} d$.  These scales, 
especially the peak spacing $\ell_A$ based on $k_A$, provide
angular size distance measures of the curvature.$^{\HW,\KSS}$  
Combined, they allow one to reconstruct the other fundamental 
cosmological parameters through their effect on the various 
physical scales (\eg\ $k_A$,$k_{\rm eq}$,$k_D$)
associated with the temperature fluctuations (see Box 2).
  
\goodbreak\bigskip
\noindent{\twelvebf Secondary Anisotropies}

\noindent
Intervening effects between recombination and the present can alter
the anisotropy.  These divide basically into two categories: gravitational
effects from metric distortions and rescattering effects from 
reionization.   Both leave imprints of
the more recent evolution of the universe and the structure within
it.  Compared with the primary signal, 
secondary anisotropies provide more details
on the evolution of structure and less robust constraints on the
background parameters.

\topinsert
\centerline{
\epsfxsize=5.5in \epsfbox{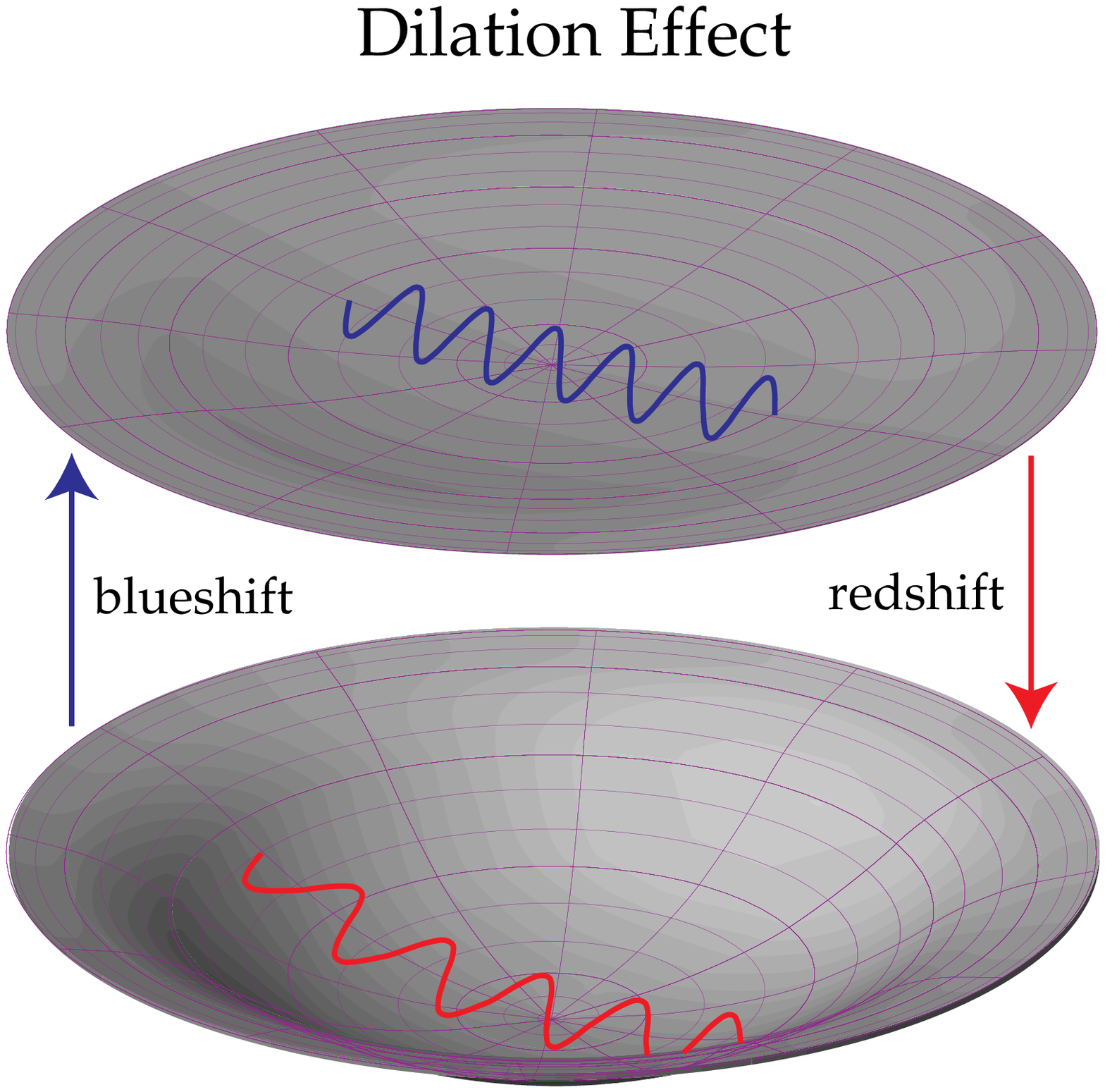}}

\baselineskip=12truept \rightskip=3truepc \leftskip=3truepc
\noindent {\bf Figure 3.} 
Dilation effect.  Changes in the spatial metric distort the wavelength
of passing photons.  If the potential decays, the contracting curvature
perturbation blueshifts the photons.  If the potential grows, the
corresponding ``stretching'' of space redshifts the photons.  Since
the curvature $\Phi \approx -\Psi$, this effect doubles the ordinary
potential redshift effect for both the driving force to the 
acoustic oscillator and the ISW effects.  
\endinsert

\noindent{\bf Gravitational Effects}

\noindent
If metric fluctuations evolve as the photons stream
past them, they leave their mark as gravitational redshifts.$^\SW$
Common manifestations include, the early ISW effect from the radiation, 
the late
ISW effect from rapid expansion, the Rees-Sciama effect from non-linear
structures, and external sources such as 
gravitational waves and topological defects.  
Metric fluctuations can also lens the photons
and distort the primary signal.

Consider first the gravitational redshift from ordinary scalar
fluctuations. 
Here the metric fluctuations
are the now familiar  
Newtonian potential $\Psi$ for the
time-time component  and the curvature perturbation $\Phi$ for
the space-space component.  As 
we have already seen, gradients in the potential $\Psi$ cause
gravitational infall and redshift.  If the depth of the potential
well changes as the photon crosses it, the blueshift from falling
in and the redshift from climbing out no longer cancel.  This leads
to a residual temperature fluctuation.  To understand what distortions
to the spatial metric through $\Phi$ do to the photons, recall
that as the universe expands,
the wavelength of the photon is also ``stretched'' through dilation 
effects.  Similarly, the overdensities which create
potential wells also stretch the space-time fabric and change
the wavelengths of the photons (see Fig.~3).  For example, as the potential
decays, the contraction of the photon wavelength blueshifts the
CMB to higher temperatures. Since $\Phi \approx -\Psi$, it 
doubles the gravitational effect on the CMB.
This heuristic explanation holds for more complicated cases 
such as vector and
tensor metric fluctuations.

\goodbreak\smallskip
\noindent
{\it Early ISW Effect}.  The early ISW effect arises if the universe
is not completely matter dominated at last scattering.$^{\HSa}$  
For adiabatic models, the potential decays 
for modes that cross the sound horizon between last scattering
and full matter domination $k_{\rm eq} \simlt k \simlt k_*$,
yielding a net effect 5 times greater than the Sachs-Wolfe 
temperature plateau
for $\Omega_0 h^2 \simlt 0.04$.  
In isocurvature models, 
potential {\it growth}
outside the sound horizon due to pressure fluctuations can contribute,
which again is larger than the Sachs-Wolfe
effect for $\Omega_0 h^2 \simlt 0.04$.   
These effects are thus
sensitive to $\Omega_0 h^2$ through the matter-radiation density
ratio $\rho_m/\rho_r$.  
The one subtlety in this and the following
effects is that it arises from a distance closer to the observer than
the primary anisotropies so that the same physical scale subtends
a larger angle on the sky (see Fig.~2).

\goodbreak\smallskip
\noindent
{\it Late ISW Effect}.
In an open or $\Lambda$  model, the universe enters
a rapid expansion phase once matter no longer dominates the 
expansion.$^{\Kofman}$  Matter-curvature equality occurs at 
$1+z_K =  \Omega_K/\Omega_0$ and matter-$\Lambda$
equality at $1+z_\Lambda = (\Omega_\Lambda/\Omega_0)^{1/3}$.  
Since the potential decays to zero within an expansion time,
independent of the wavelength,
it is the earlier of the two that matters $z_{K\Lambda} = {\rm max}
[z_K,z_\Lambda]$.
For the adiabatic case, the maximum is again 5 times greater than the
Sachs-Wolfe temperature plateau.  However 
opposing effects from decaying overdensities and underdensities
tend to cancel if the photon can travel across many wavelengths
during the decay. 
Thus, this effect is suppressed below the horizon 
at the decay epoch $k \simgt k_{K\Lambda}$.  
This scale is projected onto
a multipole moment $\ell_{K\Lambda}$ in the same manner as the primary
anisotropy.  For a fixed $\Omega_0$, 
the decay epoch occurs much later in 
flat $\Omega_K =0$  models than open 
$\Omega_\Lambda = 0$ ones.  
Thus $\Lambda$-models will suffer
cancellation of late ISW contributions 
at a much larger scale than open 
models.$^{\HSb}$
Similar agruments hold for isocurvature models, save that the late ISW
contribution exactly cancels the combined Sachs-Wolfe and early ISW
effect for $\ell \ll \ell_{K\Lambda}$.  In summary, a feature at 
$\ell_{K\Lambda}$ is expected in the CMB and can be used to 
constrain $\Omega_\Lambda$ in a flat universe or the curvature 
$\Omega_K$ of
an open universe.$^{\SS}$

\goodbreak\smallskip
\noindent
{\it Rees-Sciama Effect}.  Once fluctuations leave the linear regime,
their subsequent evolution can also make the potentials
vary with time.$^{\RS}$  In hierarchical models, where the smallest scales
go non-linear first, the effect peaks toward small scales.  
In most reasonable models, the non-linear scale
is too small to significantly affect CMB observations above the
arcminute regime.$^{\SelRS,\TL}$
Although
not
likely to be observable by the next generation of experiments,
the scale at which fluctuations become nonlinear
$\ell_{\rm NL}$ is in principle also imprinted on the CMB.

\goodbreak\smallskip
\noindent
{\it Tensors.} 
Gravitational waves introduce tensor fluctuations in the metric.  Rather
than a pure temperature shift,
these leave behind a quadrupole signature as they distort the 
distribution of passing photons.$^{\AW,\Star}$  
Gravitational
waves redshift away inside the horizon so that their main effect
on anisotropies occurs around horizon crossing.  Thus only
scales above the horizon at recombination $k \simgt k_*$ contribute
significantly. The typical
signature of gravity waves is an enhanced quadrupole $\ell=2$
and a cut off at $\ell_*$, the projected horizon at last scattering.
%These contributions are statistically uncorrelated with the
%scalar (density) contributions discussed above and thus add in 
%quadrature.  However, 
If the same mechanism generates both the
scalar and tensor fluctuations, there may exist
a relation between their spectra.  In particular, inflation 
predicts a consistency relation between the shape of the
tensor spectrum and the tensor-to-scalar amplitude ratio 
which may provide
a sensitive test of this paradigm.$^{\Davis,\KT}$

\goodbreak\smallskip
\noindent
{\it Other External Sources}.  
In general, any process that distorts the metric will produce corresponding
distortions in the CMB.  For example, topological defects may produce
a spectrum of scalar, vector and tensor perturbations.  Once generated,
scalar and tensor perturbations affect the CMB as described above.  Vector
perturbations furthermore suffer decay from the expansion and are only
important if continually generated by the source.    
Since their growth and decay is usually correlated with horizon crossing,
equal amplitude metric fluctuations as a function of scale 
at this epoch produce a roughly
scale-invariant spectrum of anisotropies 
({\it cf.}~Box 2).$^{\Coulson,\PST}$

\goodbreak\smallskip
\noindent
{\it Gravitational Lensing}.
Potential fluctuations also lens the 
CMB photons.$^{\BS,\CE}$
Gravitational lensing smears out sharp features in the spectrum of
primary anisotropy.  In a scale-invariant adiabatic 
model with $\sim 10^{-5}$
potential fluctuations initially, the smearing is less than 10\% in 
angle above 10 
arcminutes.$^{\SeljakGL}$  This reflects the fact that lensing is a second 
order effect and should be a small perturbation to the temperature
perturbation above the non-linearity scale. 

\goodbreak\bigskip
\noindent
{\bf Secondary Scattering}.

\noindent
Reionization can introduce an epoch during which the photons
are recoupled to the electrons.  Rescattering both erases primary
anisotropies and generates new secondary ones.  In models without large
small-scale fluctuations initially, such as the scale-invariant 
adiabatic model, reionization sufficiently early to make the
Compton optical depth $\tau \simgt 1$ is unlikely.  We therefore
implicitly assume that the following effects represent a 
perturbation on the primary signal. 

\goodbreak\smallskip
\noindent
{\it Rescattering Damping}.  Rescattering damps fluctuations in the
same manner as diffusion.  Scattering eliminates anisotropies leaving
them only in the unscattered fraction $e^{-\tau}$.
Since outside the horizon, fluctuations are carried by isotropic
temperature fluctuations, power is only lost on scales below
the horizon at the rescattering epoch.  
Thus the damping envelope of Fig.~B2 gains a dip at this scale 
which can in principle be used to measure or contrain the
redshift of reionization.

\goodbreak\smallskip
\noindent
{\it Doppler Effect}.  Diffusion and rescattering prevents the 
appearance of large temperature fluctuations.  However, the Doppler
effect from scattering off electrons caught in the gravitational
instability of the baryons can regenerate anisotropies.$^{\SZ}$  These
contributions are also suppressed in the same way as the late ISW effect:
photons that last scattered off overdensities and underdensities
have Doppler shifts that tend to cancel.$^{\Kaiser}$
If the underlying primary
signal is known, the angular extent of the new fluctuations
measures the horizon size at the rescattering epoch and their amplitude
probes the baryon velocity at that time. 

\goodbreak\smallskip
\noindent
{\it Non-Linear Effects}.  At very small scales, higher
order contributions are more efficient than the Doppler effect in 
regenerating anisotropies.  These generally make use of combining
the Doppler effect with variations in the optical depth.  Possibilities
include linear perturbations
in the baryon density (the Vishniac effect$^{\OV,\Vishniac}$),
inhomogeneities in the ionization fraction,$^{\ADPG}$ or contributions from
clusters (the kinematic Sunyaev-Zel'dovich effect$^{\SZkin}$).  Clusters
can also produce anisotropic spectral distortions due to the 
upscattering of photons
in frequency by hot electrons (thermal Sunyaev-Zel'dovich
effect$^{\SZkin}$).  
In models without high amplitude small-scale power, it is unlikely
that non-linear effects will dominate the
total anisotropy in the observable regime. 
%but precise 
%measurements may allow extraction of a wealth of information
%if the primary spectrum is well 
%characterized.$^{\HWvish,\PSCO,\DDP,\HT,\CB}$

\goodbreak\bigskip

\topinsert
\centerline{
\epsfxsize=5.5in \epsfbox{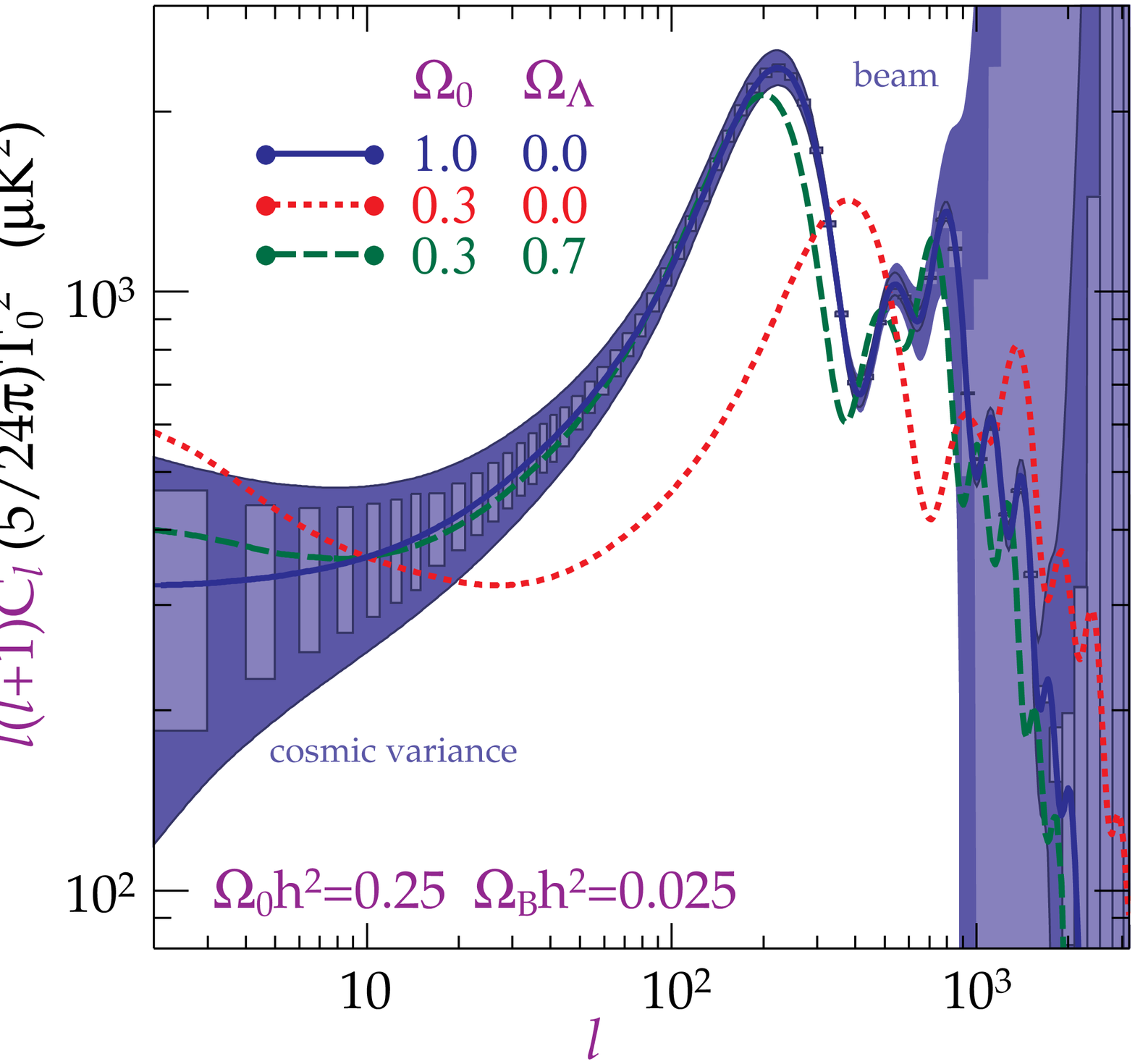}}

\baselineskip=12truept \rightskip=3truepc \leftskip=3truepc
\noindent {\bf Figure 4.} 
Idealized second ($20\mu$K sensitivity on $18'\times 18'$ pixels) 
and third
($5.5\mu$K on $10'\times 10'$ pixels) generation satellite experiments
with full sky coverage.  
For an underlying inflationary model with
$\Omega_0=1$, $\Omega_0 h^2=0.25$ and $\Omega_Bh^2=0.025$ (dark blue
curve), the power $\propto \ell(\ell+1)C_\ell$ at each multipole
$\ell \propto \theta^{-1}$ can be recovered
within the blue error contour (dark blue outline for the third
generation).  At low $\ell$, deviations
are due to ``cosmic variance,'' the same for both
experiments; at high $\ell$ from 
the beam size and sensitivity.  Binning better traces the smooth 
structure of the underlying 
power spectrum (light blue boxes $\approx 10\%$ in $\ell$).  
The $\Omega_0=0.3$ open and even the $\Lambda$ variant 
(fixed $\Omega_0 h^2$ and $\Omega_B h^2$, normalized to $\ell=10$)
are easily distinguished under the inflationary paradigm.  
More robust parameter measurements require that several acoustic features
be resolved by the third generation ({\it cf.}~Box 2).  
In practice, foreground contamination
will reduce the amount of sky available, increasing the errors as
$f_{\rm sky}^{-1/2}$, foreground subtraction will increase the noise,
and secondary anisotropies will distort the spectrum at high $\ell$.
 
\endinsert

\noindent{\twelvebf Discussion}

Primary anisotropy formation is a simple, well-understood 
linear process.  The CMB anisotropy spectrum thus provides some 
of the cleanest and most robust cosmological tests available. 
Extraction of cosmological parameters from the primary fluctuations 
requires a specific model, but the number of unverified underlying 
assumptions is greatly reduced as the resolution 
reaches the several arcminute range.  
For example, the spacing between the peaks 
is a relatively 
robust probe of space curvature. It is determined from the sound
horizon by 
the angular size-comoving distance relation and 
provides what is potentially 
the most sensitive and direct way to measure the geometry of the universe
(see Fig.~4).

Obtaining other parameters will pose a greater challenge.
In inflationary, {\it i.e.}~near scale invariant adiabatic models, 
the morphology of the first peak is controlled
by potential decay and baryon drag, which translate into additional
dependences on $\Omega_0 h^2$  and $\Omega_B h^2$.
The real advance will come with mapping of the higher 
peaks, where these various parameters have contrasting signatures 
on the peak heights and assumptions about the specific model
may be relaxed. 

To achieve these goals,
coverage of a significant part of the sky is  crucial.  
One needs   a large number 
of independent patches at any specified angular scale to 
reduce the sampling variance.  
The ultimate or so-called ``cosmic variance'' limit is 
reached with a full sky map (see Fig.~4) although in practice
some fraction of the sky will have to be discarded due to 
galactic foregrounds.
Perhaps the greatest uncertainty lying ahead will be that of 
foregrounds due to unresolved sources. 
If at least half of the sky is mapped and
relatively foreground free, one should be able
to measure most cosmological parameters presently
being considered to a precision of better than a few percent
with 10-20 arcminute resolution at a sensitivity currently
attainable. 

\bigskip
\noindent{\twelvebf Acknowledgments}

\noindent
We would like to thank E. Bunn, D. Scott, and M. White for useful discussions.
W.H was supported by the NSF and WM Keck Foundation. 
\eject

\eject

\topinsert
\centerline{
\epsfxsize=6.5in \epsfbox{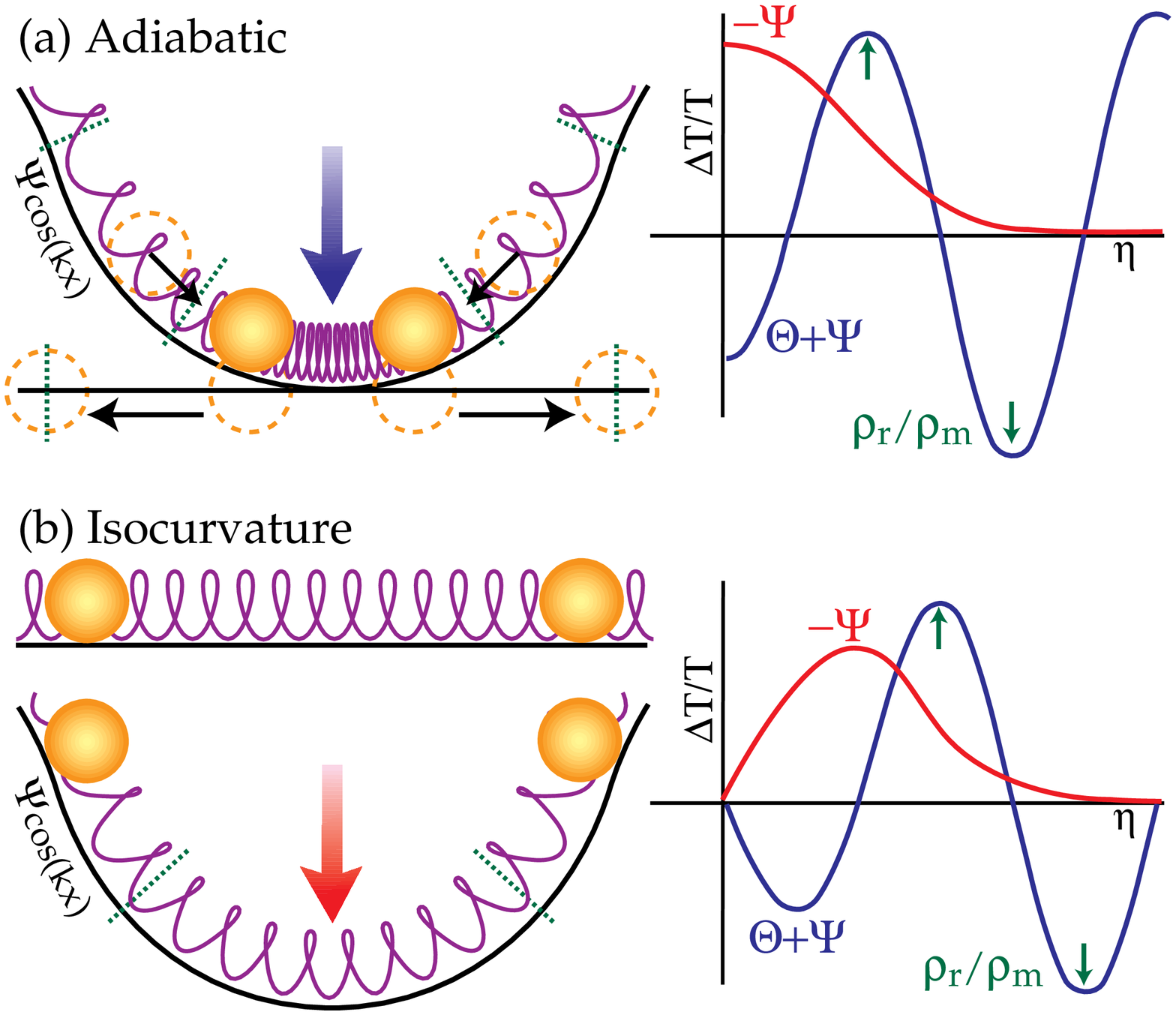}}

\baselineskip=12truept \rightskip=3truepc \leftskip=3truepc
\noindent {\bf Figure B1.}  Driven oscillations.  (a) Adiabatic
case.  As the fluid begins to compress, photon pressure
resists the increase in the density perturbation thereby allowing
the potential to decay.  Left in a highly compressed state,
the fluid then oscillates with enhanced amplitude.
(b) Isocurvature case.  Radiation fluctuations 
are set up to eliminate the potential initially. 
Photon pressure resists the accompanying rarefaction.  The fluid then
falls back into
the well, lending its self-gravity to enhance the depth.
At the compressional maxima, photon pressure again causes
potential decay leaving the
fluid in a highly compressed state.  These feedback 
effects increase with the radiation-matter density 
ratio $\rho_r/\rho_m$ driving a 
cosine oscillation in the adiabatic case and a sine oscillation
in the isocurvature case.

\endinsert

\noindent
{\twelvebf Box 1: Adiabatic vs. Isocurvature Modes}

\noindent
Adiabatic and isocurvature models differ in their initial conditions.
In the adiabatic case, number densities $n$ of the 
different particle components fluctuate together.  The 
resultant energy density perturbation $\delta$ generates 
a non-vanishing curvature fluctuation $\Phi = {1 \over 2}\delta$ 
and a gravitational potential $\Psi \approx -\Phi$.  
If radiation dominates, $\delta = \delta_r = 4\Thetan$
whereas if matter dominates $\delta = \delta_m = \delta n_r /n_r
= 3\Thetan$.  On the other hand, by balancing 
the initial fluctuations so that energy density perturbations cancel,
isocurvature fluctuations may be established.   Causality implies
that the curvature 
perturbation remains constant until 
the matter can be redistributed.$^{\HW}$  Thus
in an isocurvature model, the gravitational perturbations only
grow to be significant near horizon crossing.  Since they
represent the force felt by the oscillator, adiabatic
and isocurvature conditions yield distinctly different acoustic
signatures.  The details of how the isocurvature fluctuation is 
set up, {\it e.g.}~through balancing radiation fluctuations
with baryons, cold dark matter,
or defects such as textures,$^{\CT}$ 
is not as important as the fact that potential fluctuations vanish
initially and then grow in anticorrelation with the
photon fluctuations until horizon crossing.$^{\HW}$

In both cases, the self-gravity associated with the acoustic perturbation
can feed back into the potentials to drive the oscillator.
For the adiabatic mode, the initial potential perturbation is formed
in large part by the photon-baryon density fluctuation.  
At horizon crossing, the photon-baryon
fluid begins to compress itself due to its
self-gravity and the effective temperature reverses sign.  As pressure
tries to stop the compression, the potential decays.  The
fluid is left in a highly compressed state.  Thus self-gravity acts
as a driving term timed to enhance the first compression of
a $\cos(ks)$ series (see Fig.~B1).  This effect is doubled by dilation from
the decaying curvature perturbation $\Phi$ (see Fig.~3). 
If self-gravity dominates, as in the case in which the universe
was radiation dominated at horizon crossing, the fluid is left in 
a cosine oscillation with $2\Psi - {1 \over 3}\Psi = {5 \over 3} \Psi$
or 5 times the amplitude of the Sachs-Wolfe
tail.   Peaks occur at $mk_A = m\pi/s_*$ and baryon drag enhances
all odd peaks.

In the isocurvature case, 
photon density fluctuations are initially balanced
by those of the other species to eliminate the curvature.  
Photon pressure intercedes near horizon crossing to break this balance
letting the potential fluctuation grow from zero.  Still the photons
attempt to compensate by becoming more and more underdense or rarefied
inside potential wells $\Thetan + \Psi \approx 2\Psi < 0$ (see
also Fig.~3).  
This continues until sound horizon crossing
where photon pressure successfully resists further rarefaction 
(see Fig.~B1).
The fluid turns around and begins falling into
the potential wells and furthermore enhances them by their 
self-gravity.
As the photons resist further compression at the positive maximum,
the self-gravity contribution to the potential fluctuation decays.
This again leaves the photon-baryon fluid in a highly compressed 
state and
increases the amplitude of the acoustic oscillation.  Just as with
the adiabatic case, the self-gravity of the photon-baryon fluid essentially
{\it drives} the oscillator.  Unlike the adiabatic case, it drives
the sine rather than the cosine oscillation.   Peaks occur at
$k_m = (m-1/2)k_A$ and baryon drag enhances all even peaks.$^{\HSb}$ 
The first peak, a result of superhorizon rarefaction required
by photon compensation, is generally shallow and may be difficult
to observe.  Even so the harmonic series serves to distinguish
the two cases through the ratio of peak location to separation
$\ell_m/\ell_A = k_m/k_A$.

Isocurvature models can also possess exotic forcing
mechanisms well inside the horizon which could overwhelm the signal
described above.  
A rapidly varying potential can produce complicated acoustic
spectra, and stochastic perturbations can entirely wash out peaks
in the anisotropy spectrum.$^{\Albrecht}$ 
However, gravitational potentials inside the horizon are difficult
to generate and require extreme non-linear conditions.  
\eject
\topinsert
\centerline{
\epsfxsize=5.75in \epsfbox{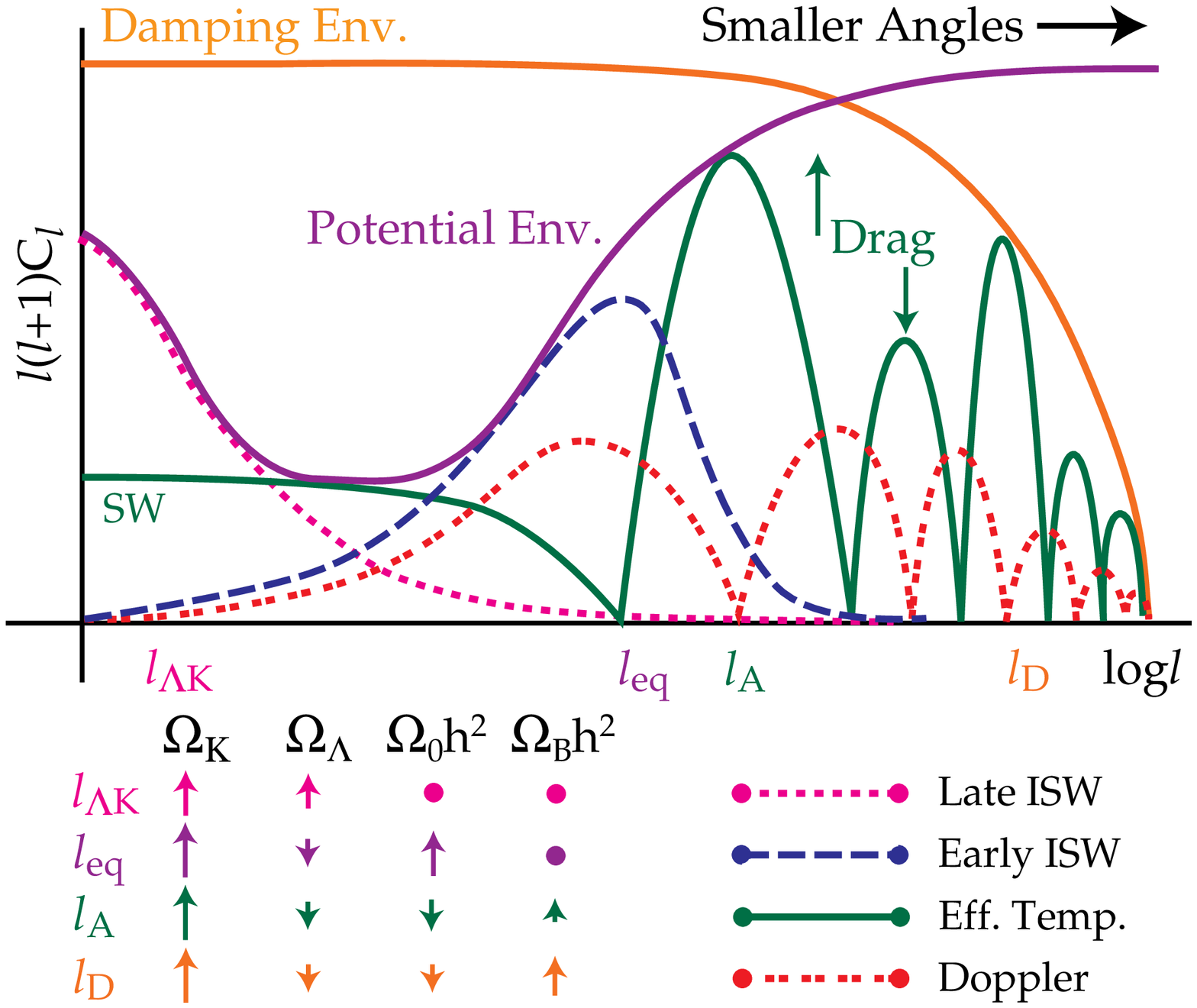}}

\baselineskip=12truept \rightskip=3truepc \leftskip=3truepc
\noindent {\bf Figure B2.} 
Anisotropy spectrum: power in anisotropies $\ell(\ell+1)C_\ell$
per logarithmic interval in $\ell \sim \theta^{-1}$.  The decomposition
into various physical effects allows one to extract four fundamental
scales in the spectrum: $\ell_{\Lambda K}$ and $\ell_{\rm eq}$ which
enclose the Sachs-Wolfe (SW) plateau in the 
potential envelope, $\ell_A$ the acoustic spacing,
and $\ell_D$ the characteristic scale of the diffusion damping envelope.
Here a scale-invariant adiabatic case is shown for illustration 
purposes. These scales may be combined to infer the four
fundamental cosmological parameters $\Omega_K (\equiv 1-\Omega_\Lambda
-\Omega_0)$, $\Omega_\Lambda$, $\Omega_0 h$
and $\Omega_B h^2$.  Baryon drag enhances all compressional (here odd) 
maxima of the acoustic oscillation, and can probe the spectrum
of fluctuations at last scattering and/or $\Omega_B h^2$.

\endinsert

\noindent 
{\twelvebf Box 2: Power Spectrum}

\noindent 
The 
scale invariant adiabatic model illustrates how the anisotropy
spectrum encodes cosmological information (see Fig.~B2).  It is
conventionally denoted $\ell(\ell+1)C_\ell$ and represents the
power per logarithmic interval 
in temperature fluctuations on angular scales $\ell \sim
\theta^{-1}$.
Highly accurate numerical results
for the spectrum in this model have long been available$^{\WS,\BE,\VS}$
with only moderate improvements to match the increasing precision
of experiments$^{\HSSW,\MB}$  (see also Fig.~4).
Here we present a more schematic description that better illuminates
the physical content and also may more easily 
be adapted to alternate models.

Several physical scales $k_{\rm feature}$ are imprinted in the
angular power spectrum
by the mechanisms described in the text.  Furthermore, each known
physical scale represents a standard ruler.  This is converted to
an angle on the sky as $\ell_{\rm feature} = k_{\rm feature} d$
by the angular diameter distance $d(K,\eta_0 -\eta_{\rm feature})$.  

The four angular scales imprinted on the CMB, $\ell_{K\Lambda}$,
$\ell_{\rm eq}$, $\ell_A$, and $\ell_D$, form the fundamental
basis for the spectrum.  The first two scales
appear because of the difference in potential evolution between
the radiation, matter, and curvature-$\Lambda$ dominated epochs.
This defines a ``potential envelope'' as depicted in Fig.~B2.  The
exact form of this envelope varies with the model
but should generally bear the imprint of the
two fundamental scales.  
For example, 
a baryon or axionic isocurvature model has an inverted form where
the maximum contributions are between $\ell_{K\Lambda}\simlt \ell
\simlt \ell_{\rm eq}$ (see Box 1), and open adiabatic models may have less
power at low $\ell$ due to a lack of super-curvature sized 
perturbations.$^{\Bucher,\YST}$

In the adiabatic case, the potential
envelope also separates cleanly into three basic pieces:
the late ISW fall, the Sachs-Wolfe (SW)
plateau, and the early-ISW/driven
oscillation rise.   Driving effects produce temperature
and sub-dominant velocity oscillations regularly spaced by $\ell_A$.
Photon diffusion
provides a damping envelope with a characteristic scale 
$\ell_D$ that is entirely independent of the initial conditions
(see Box 1).  

For the standard thermal history and a universe composed of
baryons, photons, cold dark matter, and neutrinos, these
four scales can be combined to extract the four
basic parameters$^{\HSb,\Confusion,\Seljak,\Jungman}$ 
$\Omega_K$, $\Omega_\Lambda$, $\Omega_0 h^2$, and
$\Omega_B h^2$.  Note that 
$\Omega_0 = 1 - \Omega_K - \Omega_\Lambda$ is not an independent
parameter and hence the Hubble constant $h$ may also be inferred
from these measurements. 
We choose this representation since it is
commonly supposed that either $\Omega_K$ or $\Omega_\Lambda$
vanishes and because the matter-radiation and baryon-photon
density ratios control the physical processes in the early universe.

Since $\ell_A$, the spacing between the acoustic peaks,
is only weakly dependent on the other
parameters and relies on the most distinctive features in the
spectrum, it provides the cleanest measure of the
curvature $\Omega_K$.  Even the degenerate but mild dependence
on $\Omega_\Lambda$ can be removed by measurements or constraints
on $\ell_{K\Lambda}$.  
The dramatic change in the potential
envelope at $\ell_{\rm eq}$ provides a sensitive probe of 
$\Omega_0 h^2$.  In particular $\ell_{\rm eq}/\ell_A$ is
entirely independent of $\Omega_K$ and $\Omega_\Lambda$ and
only weakly dependent on $\Omega_B h^2$ thus serving to isolate
$\Omega_0 h^2$.  
Even weaker effects such as the shift in $\ell_{\rm eq}$ and $\ell_A$ from
the neutrino mass and number are potentially observable$^{\MB,\DGS}$.
Unfortunately,
a precise measurement of $\ell_{\rm eq}$ must account for
the damping envelope.  Still, with complete information 
on the spectrum, no ambiguity arises.  Baryon drag can also 
increase the height of the first peak but its signal in the
higher peaks is unambiguous as it predicts alternating 
peak heights.  The relative peak heights measure the
baryon content times the potential at last scattering. 
To isolate $\Psi_*$, one can employ
$\ell_A/\ell_D$ which is independent of $\Omega_K$ and $\Omega_\Lambda$
and weakly dependent on $\Omega_0 h^2$ to measure the 
baryon content $\Omega_B h^2$.
Aside from gross features set by the potential envelope, the
relation between the peak locations and the peak spacing $\ell_m/\ell_A$, 
and $\Psi_*$ from baryon drag, reveal the most valuable information about
the model for the fluctuations that formed structure in the
universe (see Box 1).

As discussed in the text, complications arise if other sources
between recombination and the present are important.  
The damping envelope gains
a step at the projected horizon of a late ionization epoch.
New Doppler contributions can mask the damped peaks.
Tensor modes can introduce features in the spectrum at $\ell < \ell_A$.
Non-linear effects can alter the spectrum at $\ell > \ell_{\rm NL}$.
However, if these signals are subdominant, as is likely in many
models, they will not destroy our power to measure fundamental 
cosmological parameters from the gross features in the CMB.
Furthermore they may allow a more detailed reconstruction of 
the thermal history and evolution of fluctuations since $z \sim 10^3$ 
using precision measurements.
\eject

\noindent{\twelvebf References}

\refs\SSW. Scott, D., Silk, J. \& White, M. Science {\bf 268} 829-835
(1995)

\refs\Brandt.  Brandt, W.N. \etal\ Astrophys. J. {\bf 424} 1-21 (1994)

\refs\TegEfs.  Tegmark, M. \& Efstathiou, G. Mon. Not. Roy. Aston. Soc.
(in press, astroph/9511148)

\refs\Smoot. Smoot, G., \etal\ Astrophys. J. Lett. {\bf 396}, L1-L4 (1992)

\refs\PY. Peebles, P.J.E. \& Yu, J.T. Astrophys. J. {\bf 162}, 815-836
(1970)

\refs\SZ. Sunyaev, R.A. \& Zel'dovich, Ya.B. Astrophys. Space Sci. 
{\bf 7}, 3-19 (1970)

\refs\SW. Sachs, R.K. \& Wolfe, A.M. Astrophys. J. {\bf 147}, 73-90
(1967)

\refs\Silk. Silk, J.  Astrophys. J. {\bf 151}, 459-471 (1968)

\refs\Bardeen. Bardeen, J.M. Phys. Rev. D. {\bf 22}, 1882-1905 (1980)

\refs\Mukhanov. Mukhanov, V.F., Feldman, H.A., \& Brandenberger, R.H.
Phys. Rep. {\bf 215}, 203-333 (1992)

\refs\KS. Kaiser, N. \& Silk, J. Nature, {\bf 324} 529-537 (1986)

\refs\WSS. White, M., Scott, D., \& Silk, J. Ann. Rev. Astron. Astroph.
{\bf 32}, 319-370 (1994)

\refs\Bond. Bond, J.R. {\it Theory and Observations of the Cosmic
Microwave Background Radiation} (ed. Schaeffer, R.) (Elsevier,
Netherlands, in press)

%\refs\Bond.~Bond, J.R. {\it Distortions and Anisotropies of the Cosmic
%Background Radiation} (eds Unruh, W.G. and Semenoff, G.W.) 283-334
%(Reidel, Dordrecht, 1988)
\refs\HSa. Hu, W. \& Sugiyama, N. Astrophys. J. {\bf 444}, 489-506 (1995)

\refs\DZS. Doroshkevich, A.G., Zel'dovich, Ya.B., \& Sunyaev, R.A. 
Sov. Astron {\bf 22}, 523-528 (1978) 

\refs\HSb. Hu, W. \& Sugiyama, N. Phys. Rev. D. {\bf 51}, 2599-2630
(1995)

\refs\HW. Hu, W. \& White, M. Astrophys. J. (in press astro-ph/9602019)

\refs\SG. Sugiyama, N. \& Gouda, N., Prog. Theor. Phys. {\bf 88}, 803-844
(1992)

\refs\KSS. Kamionkowski, M., Spergel, D.N., \& Sugiyama, N. Astrophys. J. Lett.
{\bf 434}, L1-L4 (1994)

%\refs\Holtzman. Holtzman, J.A. Astrophys. J. Supp. {\bf 71}, 1-24 (1989)
%
%\refs\BEb. Bond, J.R. \& Efstathiou, G. Mon. Not. Roy. Astron. Soc. 
%{\bf 227}, 655-687 (1987)
%
%\refs\Harrison. Harrison, E.L. Rev. Mod. Phys. {\bf 39}, 862-882 (1967)
%
%\refs\Wilson. Wilson, M.L. Astrophys. J. {\bf 273}, 2-15 (1983)
%
%\refs\KS. Kodama, H. \& Sasaki, M. Prog. Theor. Phys. Supp. {\bf 78}, 1-166
%(1984)
%
%
%\refs\Bond.~Bond, J.R. {\it Distortions and Anisotropies of the Cosmic
%Background Radiation} (eds Unruh, W.G. and Semenoff, G.W.) 283-334
%(Reidel, Dordrecht, 1988)
%
%\refs\Jorg. J{\o}rgensen, H.E., Kotok, E., Naselsky, P., \& Novikov, I. 
%Astron. Astrophys. {\bf 294}, 639-647
%
%\refs\HSsmall. Hu, W. \& Sugiyama, N. astroph-?????.
%
%\refs\Kaiser. Kaiser, N. Mon. Not. Roy. Astron. Soc. {\bf 202}, 1169-1180
%(1983)
%
%\refs\HSSW. Hu, W., Scott, D., Sugiyama, N. \& White, M. Phys. Rev. D
%(submitted)

\refs\Kofman. Kofman, L.A. \& Starobinskii, A.A. Sov. Astron. Lett. {\bf
9}, 643-651 (1985)

\refs\SS. Sugiyama, N. \& Silk, J. Phys. Rev. Lett. {\bf 73}, 509-513
(1994)

\refs\RS. Rees, M.J. \& Sciama, D.N. Nature {\bf 519}, 611-? (1968)

\refs\SelRS. Seljak, U. Astrophys. J. {\bf 460}, 549-555 (1996)

\refs\TL. Tuluie, R. \& Laguna, P., Astrophys. J. Lett. {\bf 445}, L73-L76
 
%\refs\DNP. Doroshkevich, A.G., Novikov, I.D., \& Polnarev, A.G.,
%	Sov. Astron. {\bf 21}, 529-535 (1977)

\refs\AW. Abbott, L.F. \& Wise, M.B. Nucl. Phys. {\bf B244}, 541-548 (1984)

\refs\Star. Starobinskii, A.A., Sov. Astron. Lett. {\bf 11}, 113 (1985)

%\refs\Crittenden. Crittenden, R. \etal\ Phys. Rev. Lett. {\bf 71}, 324-327
%(1994)

\refs\Davis. Davis, R. \etal\ Phys. Rev. Lett. {\bf 69}, 1856-1859 (1992);
	(erratum {\bf 70}, 1733)

\refs\KT. Knox, L. \& Turner, M.S., Phys. Rev. Lett. {\bf 73}, 3347-3350 
	(1994)

%\refs\BSB. Bennett, D.P., Stebbins, A., \& Bouchet, F.R. Astrophys. J.
%Lett.  {\bf 399}, L5-L8. (1992)

\refs\Coulson. Coulson, D., Ferreira, P., Graham, P., \& Turok, N. Nature
{\bf 368}, 27-31 (1994)

\refs\PST. Pen, U.L., Spergel, D.N., \& Turok, N., Phys. Rev. D. {\bf 49},
692-729 (1994)

\refs\BS. Blanchard, A. \& Schneider, J. Astron. Astrophys. {\bf 184}, 1-6
(1987)

\refs\CE. Cole, S. \& Efstathiou, G. Mon. Not. Roy. Astron. Soc. {\bf
239}, 195-200 (1989)

\refs\SeljakGL. Seljak, U. Astrophys. J. {\bf 463} 1  (1996)

\refs\Kaiser. Kaiser, N. Astrophys. J. {\bf 282}, 374-381 (1984)

\refs\OV. Ostriker, J.P. \& Vishniac, E.T. Astrophys. J. Lett. {\bf 306}, 
L51-L58 (1986)

\refs\Vishniac. Vishniac, E.T. Astrophys. J. {\bf 322}, 597-604 (1987)

\refs\ADPG. Aghanim, N., Desert, F.X., Puget, J.L., \& Gispert, R. 
Astron. \& Astrophys (in press astro-ph/9604083)

\refs\SZkin. Sunyaev, R.A. \& Zel'dovich, Ya. B. Comm. Astrophys. 
Space Phys. {\bf 4}, 173-178 (1972)

%\refs\HWvish. Hu, W. \& White, M. astro-ph/9507060 (1995)
%
%\refs\PSCO. Persi, F.M., Spergel, D.N., Cen, R. \& Ostriker, J.P.
%Astrophys.~J. {\bf 442} 1 (1995)
%
%\refs\DDP. De Luca, A., Desert, F.X., Puget, J.L. Astron. \& Astrophys
%{\bf 300} 335 (1995)
%
%\refs\HT. Haenhelt, M.G. \& Tegmark, M. astro-ph/9507077
%
%\refs\CB. Ceballos, M.T. \& Barcons, X., Mon. Not. Roy. Astr. Soc.
%{\bf 271}, 817-826 (1994)

%\refs\KSY. Kawasaki, M., Sugiyama, N. \& Yanagida, T. hep-ph/9512368
%(1995)

\refs\CT. Crittenden, R. \& Turok, N. Phys. Rev. Lett. {\bf 75} 2642-2645
(1995)

\refs\Albrecht. Albrecht, A., Coulson, D., Ferreira, P. \& Magueijo,
J. Phys. Rev. Lett. {\bf 76} 1413-1416 (1996)

\refs\WS.  Wilson, M. \& Silk, J. Astrophys. J. {\bf 243}, 14-25 (1981)

\refs\BE. Bond, J.R. \& Efstathiou, G. Astrophys. J. Lett {\bf 285}, L45-L48
(1984)

\refs\VS. Vittorio, N. \& Silk, J. Astrophys. J. Lett {\bf 285}, 
L39-L43 (1984)

\refs\HSSW. Hu, W., Scott, D., Sugiyama, N. \& Silk, J. Phys. Rev. D
{\bf 52} 5498-5515

\refs\MB. Ma, C-P. \& Bertschinger, E. Astrophys. J. {\bf 455}, 7-25 (1995)

\refs\Confusion. Bond, J.R. \etal\ Phys. Rev. Lett. {\bf 72}, 13-16 (1994)

\refs\Seljak. Seljak, U. Astrophys. J. Lett. {\bf 435}, L87-L90 (1994)

%\refs\LS. Lyth, D.H. \& Stewart, E.D.  Phys. Lett. B {\bf 252}, 336-353 (1990)

\refs\Jungman. Jungman, G., Kamionkowski, M., Kosowski, A. \& Spergel, 
astro-ph/9512139 (1995)

%\refs\RP. Ratra, B. \&  Peebles, P.J.E. Astrophys. J. Lett. {\bf 432}, L5-L9
%(1994)

\refs\Bucher. Bucher, M., Goldhaber, A.S., \& Turok, N. Phys. Rev. D
{\bf 52}, 3314-3337 (1995)

\refs\YST. Yamamoto, K., Sasaki, M. \& Tanaka, T. Astrophys. J. {\bf 455},
412 (1995)

% \refs\LW. Lyth, D.H \&  Woszczyna, A, astroph-9501044.
%
%\refs\TWL. Turner, M.S., White, M., \& Lidsey, J.E. Phys. Rev. D {\bf 48},
%4613-4622 (1993)

\refs\DGS. Dodelson, S., Gates, E. \& Stebbins, A., astro-ph/9509147
\end